\documentclass[showpacs, twocolumn]{revtex4-1}
\usepackage{graphicx}
\usepackage{amsmath}
\usepackage{csquotes}

\begin{document}

\title{ Interwires polar soliton molecules in a biwire system}

\author{Khelifa Mohammed Elhadj$^{1,2}$, Abdel\^{a}ali Boudjem\^{a}a$^{1,2}$ and U. Al-Khawaja$^{3}$}

\affiliation{$^1$Department of Physics, Faculty of Exact Sciences and Informatics, 
and $^2$Laboratory of Mechanics and Energy, Hassiba Benbouali University of Chlef, P.O. Box 78, 02000, Chlef, Algeria.
\\$^3$Physics Department, United Arab Emirates University, P.O. Box 15551, Al-Ain, United Arab Emirates.} 


\date{\today}

\begin{abstract}

We investigate stability the dynamical properties of one-dimensional interwires polar soliton molecules in a biwire setup 
with dipole moments aligned perpendicularly to the line of motion and in opposite directions in different wires.
Numercial results based on a nonlocal nonlinear Schr\"odinger  equation reveal the existence of a bound state leading
to the formation of a molecule. 
We show that the degree of the nonlocality and the interwire separation play an important role in stabilizing the molecules.
Dynamics of the width and center-of-mass of a solitons is also analyzed in terms of the system parameters.

\end{abstract}


\maketitle

\section{Introduction} 

Matter-wave solitons in dipolar Bose-Einstein condensates (BECs) have attracted much attention in the past decade owing to their fascinating attributes,
see for review \cite{Baranov, Lay}.
The nonlocality  originating from the dipole-dipole interactions (DDI) may significantly affect the physics of solitons.
It provokes a nonlocal character to soliton-soliton interactions \cite {Ped, Cuev, Bland, Pawl, Edm}.
The competition between local and nonlocal interactions leads to the formation of soliton molecules of bright \cite {Cuev, Lak} and dark solitons \cite{Pawl, Baiz}.
The stability and the dynamics of both bright and dark solitons in quasi-one-dimensional (1D) dipolar BEC were studied in \cite {Cuev,  Bland,Pawl, Edm, Lak, Abdul, Sinh,Yong}.
In addition, discrete and vortex solitons have been analyzed in dipolar BECs \cite{Tik, Gli, Ai, Adh, Fan}.
It has been shown also that stable and mobile dark-in-bright solitons can be formed in dipolar binary BEC \cite {Adh1}.
On the other hand, anisotropic solitons have been predicted in dipolar BECs \cite {Ped, Edm, Tik1, Eic, Kob, Rag} due to the anisotropy of the DDI.
Unlike the short-range interacting condensates,  quasi-2D and 3D dipolar BECs can support the 
emergence of stable solitons \cite {Ped, Nath, Tik1, Eic, Rag}, where the famous “snake” instability of dark solitons is suppressed \cite{Nath}.

Recently, quantum degenerate dipolar gases placed in equidistant layers/wires have recieved a great deal of interest.
These layered/wired structures which are connected with each other due to the long-range character of the DDI, 
exhibit remarkable interlayer effects such as interlayer  bound states and superfluids of polar molecules \cite {Wang, Piko, Misha,  Shi, Pot, Ros, Vol, Dalm}.
In the nonlinear physics, the interlayer/interwire interaction leads to an effective potential between separated solitons, 
result in the formation of soliton molecules \cite{Santos1, Santos2}.
Such molecules vitally depend on the interplay between contact and dipolar interactions \cite{Rag, Santos1, Santos2}.

\begin{figure}
\centerline{
\includegraphics[scale=0.6]{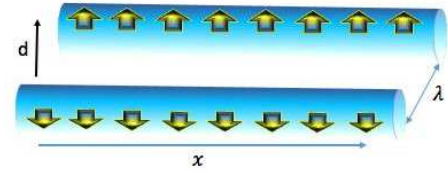}}
 \caption{ Biwire system of cold polar molecules with dipoles oriented in opposite directions in different wires.}
 \label{schm}
\end{figure}

In this paper we consider dipolar BECs loaded in a biwire (tubes) system of a quasi-1D trap, 
where the two wires are separated by a distance $\lambda$ much larger than the confinement length
of the molecules within each wire ensuring vanishing hopping between wires. 
The dipole moments are assumed to be aligned perpendicularly to the tubes by an external field and in opposite directions in different wires (see Fig.\ref {schm}). 
The long-range nature of the DDI affords an interaction in one BEC inside each wire. 
The most important feature of this setup is that the anisotropy of the DDI connects particles from each wires in a very specific form (different from the intrawire interaction) 
leading to a two-component BEC \cite{Misha}.
Experimentally, this biwire geometry can be created by optical lattices or atomic chip traps \cite{Chen}.

In contrast to the arrangement used in \cite {Wang, Piko, Misha,  Shi, Pot, Ros, Vol},
our biwire configuration makes the interwire interaction to be repulsive at short distance regime and attractive in the long-ranged regime.
For repulsive contact interactions, it may open a new route to explore dark soliton molecules due to the special interwire effects 
not considered in the configuration of \cite {Wang, Piko, Misha,  Shi, Pot, Ros, Vol}. 
This biwire pattern can also lead to the formation of matter-wave bright soliton molecules for  both short- and long-range attractive interactions.
In quasi-2D geometry, this type  of system supports the formation of unconventional interlayer superfluids 
such as extended $s$-, $p$-, $d$-, and $f$-wave superfluids \cite{Fedo, Boudj, Boudj1}.

The numerical simulation of the underlying nonlocal nonlinear Schr\"odinger  equation, 
reveals that the stability and properties of individual solitons, and soliton molecules in such a geometry crucially depend on the interwire distance and on 
the interplay between contact and dipolar interactions.  
We show that in such a biwire system, the excitation spectrum can acquire a roton-maxon structure.
A detailed analysis of the nonlinear dynamics of these exotic soliton molecules is also carried out.  
It is found that near the roton instability, the solitons complexes exhibit oscillatory behavior with increasing the interaction stength and the interwire distance. 


\section{Model} \label{model}

Let us consider the quasi-1D biwire configuration of Fig.\ref {schm}, where at each wire a dipolar BEC of $N$ atoms 
with dipoles are head-to-tail across the wires.
At zero temperature, the system is governed by the following dimensionless nonlocal  two-coupled  GP equations 
\begin{align}  \label{NLSE}
i \frac{ \partial \psi_j}{ \partial t} &=- \frac{ \partial^2 \psi_j } {\partial x^2} - g |\psi_j|^2 \psi_j\\
&+\int dx' V_d (|x-x'|) \left(|\psi_1 (x')|^2 + |\psi_{-1}(x')|^2\right)  \psi_j,  \nonumber
\end{align}
where $j=\pm 1$ is the wire index, $\psi_j$ is the wavefunction at each wire which is normalized to the number of
atoms in the condensate $\int |\psi_j|^2 dx=N$, and $g=2a/l_0$  is the strength of the local nonlinearity with
$l_0=\sqrt{\hbar/m \omega_{\perp}}$ being  the harmonic oscillator length, $\omega_{\perp}$ corresponds to the frequency of radial confinement,
and $m$ is the particle mass.
In system (\ref{NLSE}), lengths, time, density and energies are expressed in units of $l_0$, $\omega_{\perp}^{-1}$, $l_0^{-1}$ 
and $\hbar \omega_{\perp}$, respectively.  \\

The DDI potential $V_d (x)$  appearing in Eq.(\ref{NLSE}) is composed of two parts: the interwire interaction potential is
\begin{equation} \label{pot}
V_{\text{inter}} (x) = - g_d  \frac{x^2-2\lambda^2}{(x^2+\lambda^2)^{5/2}},
\end{equation}
where $g_d= 2 r_*/ l_0$  with $r_*=md^2/\hbar^2$ being the characteristic dipole-dipole distance,
and $d$ is the scalar product of the average dipole moments of the molecules.
The potential $V_{\text{inter}} (x)$ is repulsive for $x < \sqrt{2} \lambda$, while it is attractive at  $x > \sqrt{2} \lambda$
which may open up the possibility of forming interwire soliton molecules.
The soliton-soliton potential (\ref{pot}), in contrast to that used in Refs \cite{Santos1, Santos2}, is attractive at large distance $x$ 
which may lead to the formation of an interwire bound state of two solitons in different wires (soliton dimer). 
When the interwire distance is increased, the local minimum is shifted and thus, the molecular structure is modified.
On the other hand, the intrawire interaction has a complicated form \cite{Sinh, Cai}.
However the most convenient form for analytical calculation is that proposed in  Ref \cite{Cuev} 
\begin{equation} \label{pot1}
V_{\text{intra}}(x) =g_d \frac{\delta^3}{(x^2+\delta^2)^{3/2}},   
\end{equation}
where $\delta=\pi^{-1/2}$ is a cutoff parameter, which regularizes the singularity problem. \\
For $\lambda \gg l_0$, both $V_{\text{intra}}(x)$ and  $V_{\text{inter}} (x)$ depend only on
the interparticle distance along the wire direction $x$.
In the absence of intertube interactions, the dynamics of solitons and soliton molecules in each 1D tube have been extensively investigated 
using the standard nonlocal GP equation (see e.g. \cite{Cuev, Cai, Abdul, Baiz}).

The soliton's energy functional corresponding to Eq.(\ref{NLSE}) reads
\begin{align}\label{SEngy}
E=&\sum_j \bigg\{ \frac{1}{2} |\nabla \psi_j|^2-\frac{g}{2} \bigg ( |\psi_j|^4  +\epsilon_{dd} |\psi_j|^2 \\
& \times \int_{-\infty} ^{+\infty} V_d(x-x') \left[|\psi_1(x')|^2+|\psi_{-1}(x')|^2 \right] d x'  \bigg) \bigg\}  \nonumber.
\end{align}
The first term represents the kinetic energy, the second term stands for the contact interactions
energy, and the third term accounts for interaction energy due to DDI.

\section{ Stability of the uniform system}

\begin{figure}
\centerline{
\includegraphics[scale=0.455] {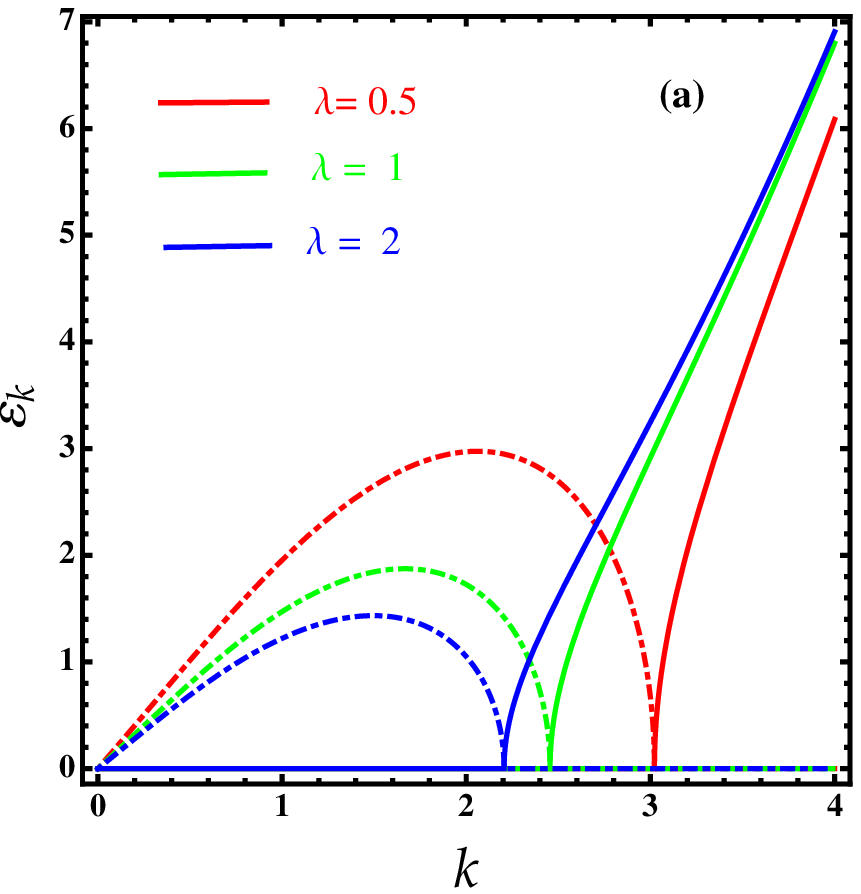}
\includegraphics[scale=0.45] {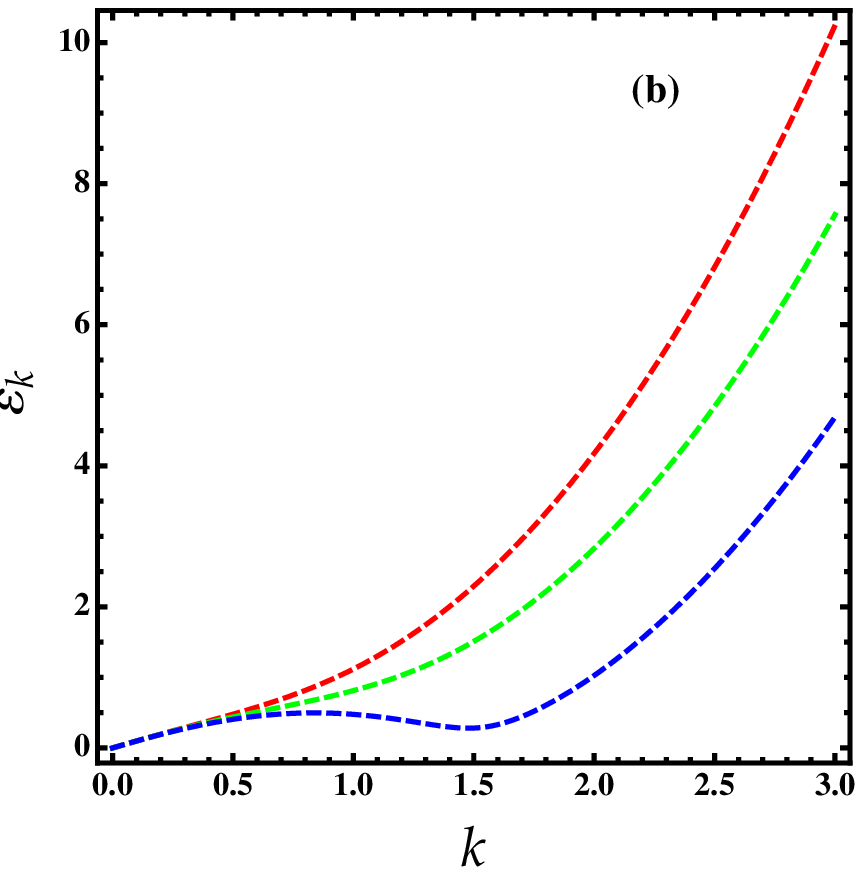}}
 \caption{(Color online) The Bogoliubov spectrum from Eq.(\ref{Bog}) for $\epsilon_{dd}=1.5$. 
(a) $\mu_0=- 1$ (solid lines are the real part and  dotdashed lines represent the imaginary part). (b) $\mu_0=1$.}
 \label{PRS}
\end{figure}

Here we analyze the stability of the interwires nonlinear modes associated with Eq.(\ref{NLSE}) which is
affected by the excitations of the condensate.
The elementary excitation energies $\varepsilon_k$ can be obtained by considering small perturbations 
of the order parameter around the equilibrium solution $\psi_{0j}$ namely:  
$\psi_j=\psi_{0j}+ \delta \psi_j$, where $ \delta \psi_j \ll \psi_{0j}$. This is the well known Bogoliubov theory \cite {Baranov, Lay, PitaevString, Boudj4}.
In the homogeneous case, the Fourier transform of the potential (\ref{pot}) reads $V_{\text{inter}} (k) = g_d k^2 [K_0( k\lambda)+K_2( k\lambda)]$,
where $K_0(y)$ and $K_2(y)$ are modified Bessel functions, and the small amplitude fluctuations are given as
\begin{equation} \label{pert}
\delta \psi_j= \left(u_k e^{-ik x+i \varepsilon_k t}+ v_k e^{ik x- i \varepsilon_k t} \right) e^{-i\mu t},
\end{equation}
where $\mu$ is the chemical potential and $u_k$, $v_k$ are the quasi-particle amplitudes. 
The Bogoliubov dispersion relation turns out to be given as
\begin{equation} \label{Bog}
\varepsilon_k= k \sqrt{ \frac{k^2}{4}+ \mu_0 \left[ 1+ \epsilon_{dd} k^2 \left( K_0( k\lambda)+K_2( k\lambda) \right) \right] },
\end{equation}
where $\epsilon_{dd}=g_d/g$ is the relative strength  describing the interplay between the DDI and short-range interactions and $\mu_0=g \psi_0^2$.
Based on the sign of $g$ and the strength of the DDI, the system sustains two types of instabilities namely, phonon instability (PI) and roton instability (RI).
In the limit of low momenta ($k \rightarrow 0$), the PI occurs for $ g <0$ whatever the value of $\lambda$ as is shown in Fig.\ref{PRS}.a. 
In nondipolar homogeneous BECs, the PI arises only for $ g <0$ (see e.g. \cite{Don}), 
while  for BEC-impurity mixtures it originates even for repulsive short-range interaction \cite {Boudj2}.
In the case of 3D dipolar BECs, the PI comes from the anisotropy of DDI, resulting in a local collapse, see for review \cite{Lay}.

For $ g >0$ and under certain critical value of the interwire distance,  
the dipoles can modify the form of the dispersion relation exhibiting first a maximum and then a minimum in spectrum as the momentum increases results in 
roton-maxon structure as is seen in Fig.\ref{PRS}.b.
If the roton minimum touches the zero-energy axis, the Bose condensate suffers a RI. 
In this case, the energy and the position of the roton can be modified by changing $\epsilon_{dd}$.
The same behavior holds in quasi-2D dipolar BECs \cite {Rag, Boudj3}. 
The height of the roton is sensitive to $\lambda$; for sufficiently large intrawire space, 
the roton approaches zero which destroys the dimerization of solitons modeled initially by two dipolar BECs placed in a biwire system.

\section{ Dimer soliton molecules}\label{DSM}

In this section we analyze the rudiments of dimer soliton molecules in biwire systems. 
The results presented here rely on numerical simulation of the nonlocal GP equation (\ref{NLSE}), 
using a suitable initial condition envisaging solitonic solutions. 

The potential of interaction between the solitons can be calculated numerically from the trajectories of the solitons. 
The separation is calculated then differentiated numerically twice. 
This will give the force. The potential is then obtained by integrating the force with respect to the separation.


\begin{figure}
\centerline{
\includegraphics[scale=0.8]{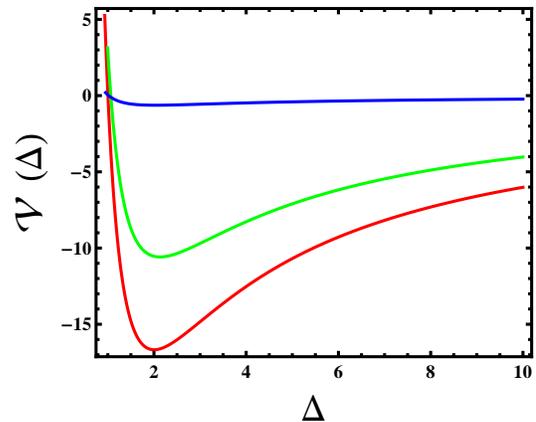}}
 \caption{ (Color online) Soliton-soliton potential as  a function of sepration $\Delta$ from numerical simulation of the nonlocal GP equation (\ref{NLSE}) for several values of $\lambda$.  
Parameters are: $g=7.9$,  $g_d=0.4$. 
Red line: $\lambda=0.5$. Green line: $\lambda=1$. Blue line: $\lambda=2$.}
 \label{LocM}
\end{figure}

Figure \ref{LocM} shows that the soliton-soliton potential ${\cal V}$ predicted by our numerical simulation 
has a local minimum at $\Delta= 2$ indicating the formation of a molecule of dimer solitons. 
The depth of the effective potential minimum is decreasing with increasing the interwire distance $\lambda$ i.e the two solitons become less bounded 
and hence, the molecule undergoes instability.  

\begin{figure}
\centerline{
\includegraphics[scale=0.8]{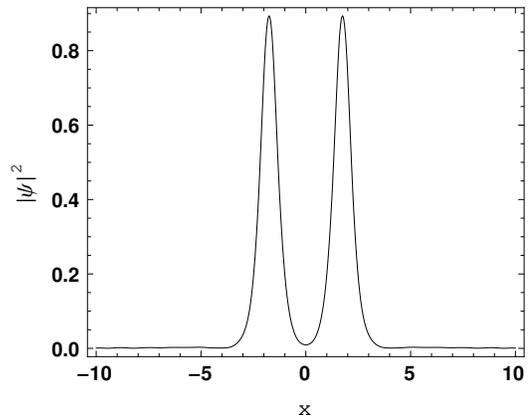}}
 \caption{ Stationary localized solutions of the nonlocal GP equations (\ref{NLSE}). 
 Parameters are $g=7.9$,  $g_d=0.4$ and $\lambda=1$.}
 \label{PSS}
\end{figure}

Figure \ref{PSS} displays the shape of a two-soliton molecule obtained by numercially solving the static nonlocal GP equations (\ref{NLSE}).
As is seen the two solitons are stable and symmetric.

\begin{figure}
\centerline{
\includegraphics[width=2.8 cm,height=5 cm]{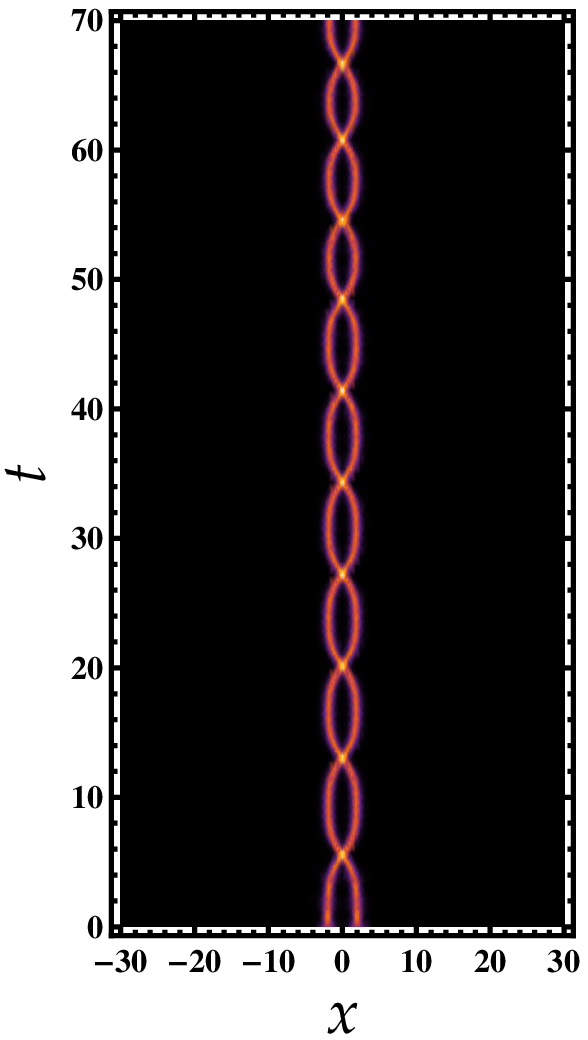}
\includegraphics[width=2.8 cm,height=5 cm]{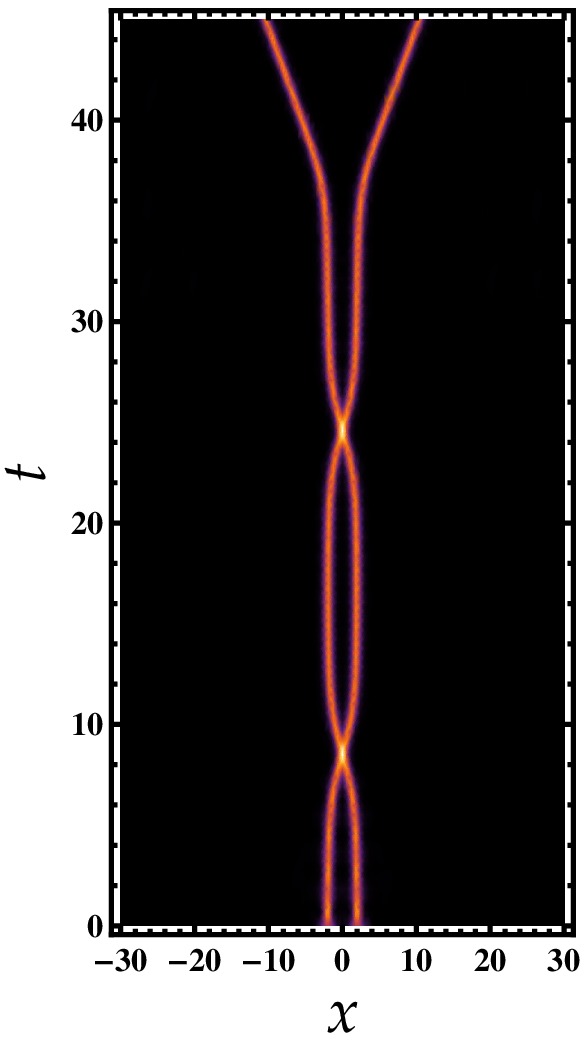}
\includegraphics[width=2.8 cm,height=5 cm]{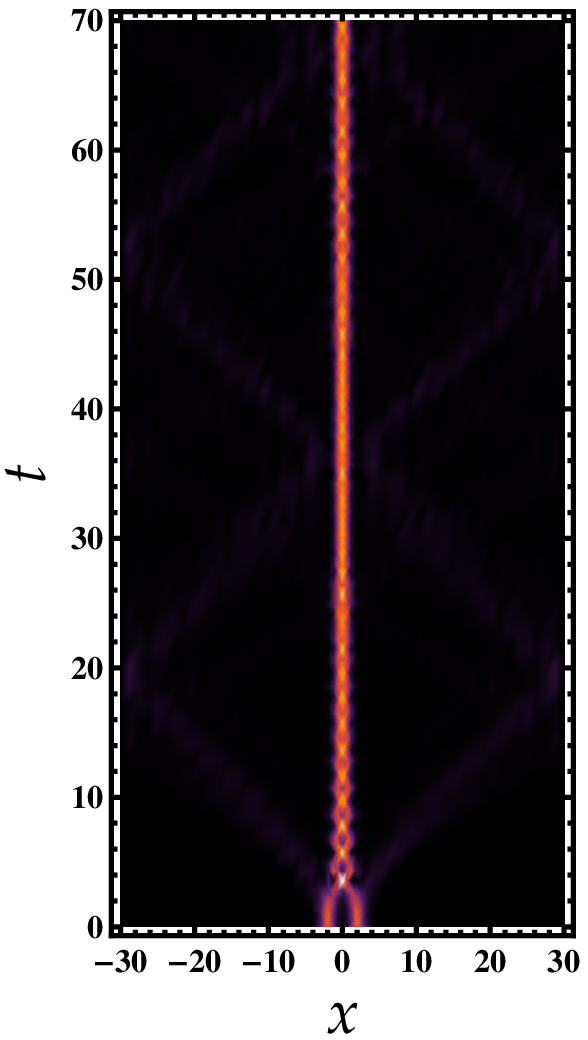}}
 \caption{ (Color online) Spatiotemporal evolution of the interwire dimer soliton molecule. 
(left) For $\lambda =1$ the molecule is stable.
(middle) For $\lambda =2$ the solitons repel. 
(right) For $\lambda =0.5$ molecule vibrations are clear.
Parameters are the same as in Fig.\ref{LocM}.}
\label{Dyn}
\end{figure}
Figure.\ref{Dyn} depicts the density profiles which were calculated using the equilibrium conditions found by minimizing the energy functional (\ref{SEngy}). 
For small separation, the solitons attract each other undergoing fusion and eventually becoming a single big soliton (see the right pannel).
If solitons are well separated ($\lambda >1$), they can be approximated as point-like solitons \cite{Nic}. 
In this case, the solitons repel and acquire sufficient kinetic energy to diverge to infinity at larger time as is shown in the middle pannel. 
When $\lambda =1$, the molecule exhibits periodic oscillations around the equilibrium during its propagation analogous to vibrational modes of a diatomic molecule 
(see the left pannel). 

\begin{figure}
\centerline{
\includegraphics[scale=0.45]{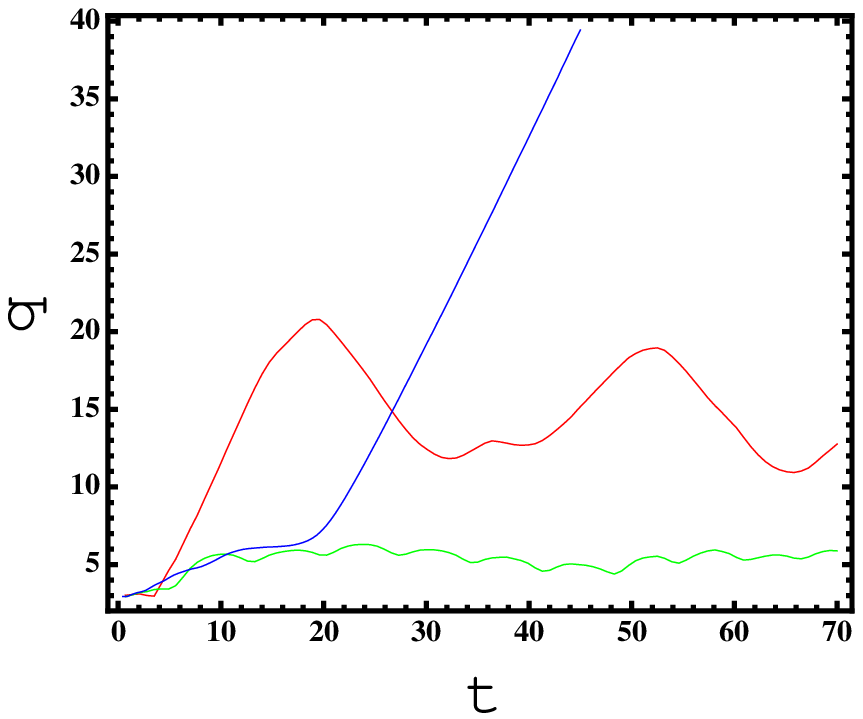}
\includegraphics[scale=0.46]{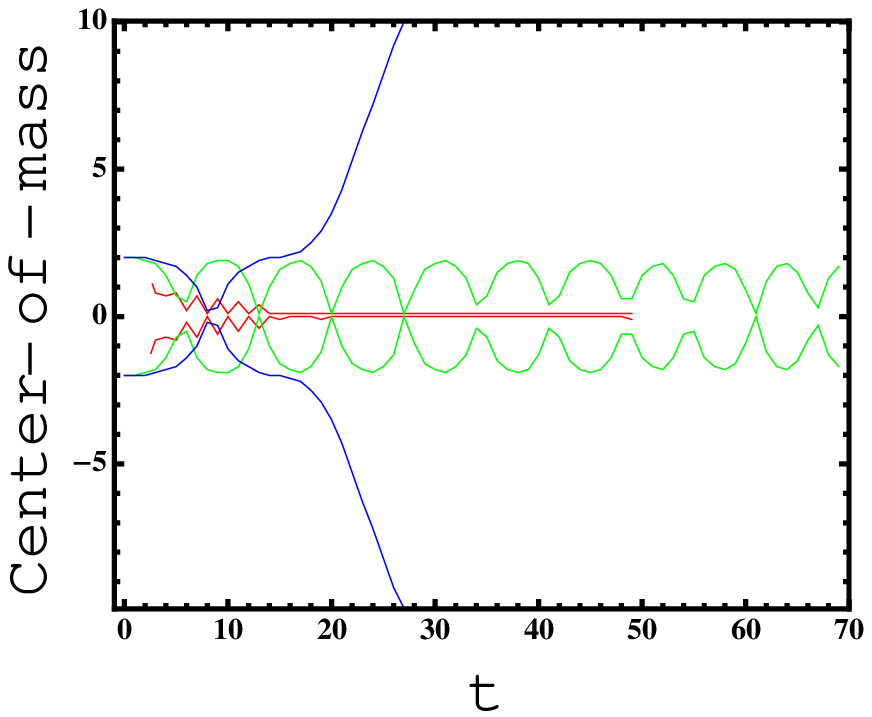}}
 \caption{ (Color online) Time evolution of the width of the soliton along $x$ direction (left panel) and   
the centers of masses of two solitons, forming the molecule placed in opposite directions (right panel). 
Red line: $\lambda=0.5$. Green line: $\lambda=1$. Blue line: $\lambda=2$. Parameters are the same as in Fig.\ref{LocM}.}
 \label{P2}
\end{figure}

Figure \ref{P2} reveals that both the width and the center-of-mass of the soliton exhibit breathing (contraction and expansion) oscillations. 
For large interwire speration $(\lambda \geq 2$), process of oscillation of the solitons goes on for some time ($t<50$) due to the roton instability
where the molecule disintegrates into individual freely solitons. 
The center-of-mass and the width of soliton can be extracted from the GP equation (\ref{NLSE}) using 
$\eta_j (t) = \int_{-\infty} ^{+\infty} d x\, x  |\psi_j(x,t)|^2$ and $q_j(t) =  \sqrt{   \int_{-\infty} ^{+\infty} d x\, x^2  |\psi_j(x,t)|^2 }$, respectively.

\begin{figure}
\centerline{
\includegraphics[scale=0.45]{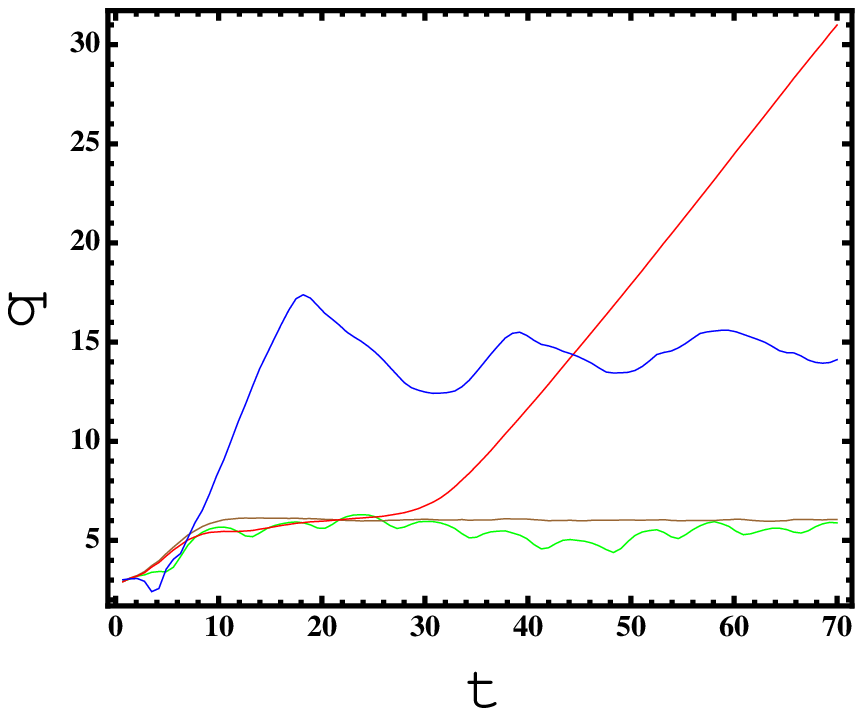}
\includegraphics[scale=0.46]{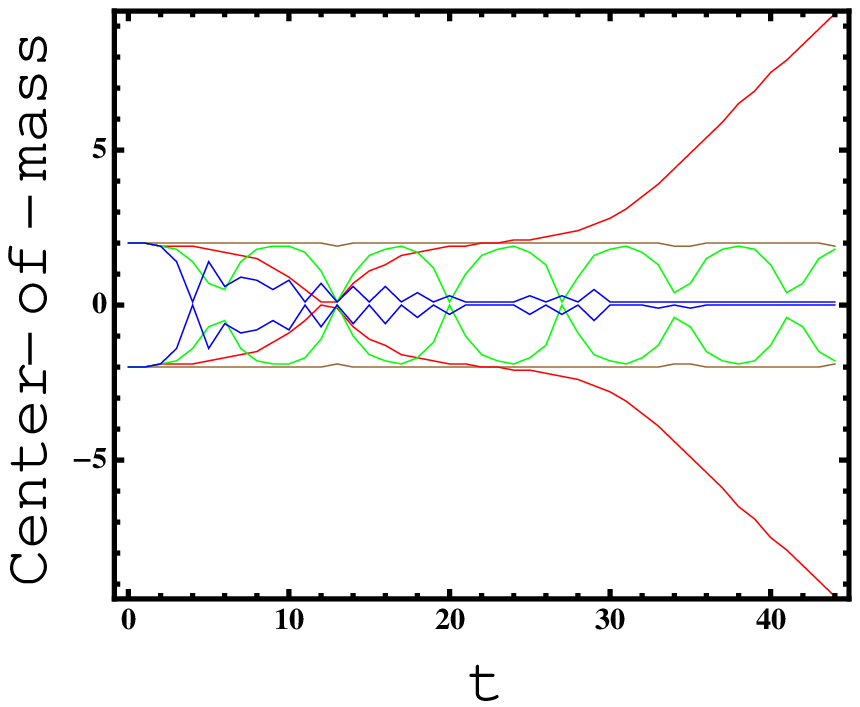}}
 \caption{ (Color online) Time evolution of the width of the soliton along $x$ direction (left panel) and   
the centers of masses of two solitons (left panel)  for different values of $g_d$. 
Brown line: $g_d=0$.  Red line:   $g_d=0.05$.  Green line: $g_d=0.4$. Blue line: $g_d=1.4$.
Parameters are $g=7.9$,  and $\lambda=1$. }
 \label{P2gd}
\end{figure}

We now examine the role of the interaction strength in the behavior of the width and the center-of-mass. 
By increasing the interaction strength ($g_d \simeq 1.4$), the two solitons become very close to each other and 
the modulation of oscillations of width and the center-of-mass of the solitons becomes stronger 
as is seen in Fig.\ref{P2gd}.   The oscillations decay at larger time. 
This can be understood from the fact that for strong interaction, the atoms scattered with high energy leave the condensate due to the absence of an external trap.
For $g_d=0.4$, the solitons continue to oscillate periodically forever. 
For $g_d=0$, the two solitons undergo very slow oscillations and the molecule cannot be formed in such a situation.

\section{Conclusion} \label{Conc}

In this paper we numerically studied interwire effects on the formation of polar soliton molecules in a biwire system 
where dipole moments are aligned head-to-tail across the wires.
We showed that the intertube interactions play a key role in the formation of a stable bound state (molecule)
of unconnected solitons.  The stability and the interaction of such dimer molecules depends also on the interwire distance. 
The breathing oscillations of the width and the center-of-mass of the soliton have been deeply analyzed
in terms of the interaction strength, rotonization effects and the interwire separation. 

The realization of such polar soliton molecules in a quasi-1D biwire system is less challenging and can be probed in current Cr, Dy and Er experiments. 
For instance, for ${}^{164}$Dy atoms which are characterized by $r_*=131 a_0$, where $a_0$ is the Bohr radii and $\omega_{\perp}=60 \times 2 \pi$ Hz
placed in a biwire system, a stable soliton molecule can be nucleated for interwire separation $\lambda \simeq 400$ nm.

Interwire effects on the generation and the stability of dark solitons in biwire systems is of great interest for future studies.

\section*{Acknowledgements}
We acknowledge support from the University of Chlef, and from UAE University through the grants UPAR(7)-2015, UPAR(4)-2016, and UPAR(6)-2017.
We thank Sadhan Adhikari for useful comments on the manuscript.


\begin{thebibliography} {28}

\bibitem{Baranov}  M. A. Baranov, Physics Reports {\bf 464}, 71 (2008).
\bibitem {Lay} T. Lahaye, C. Menotti, L. Santos, M. Lewenstein, and T. Pfau, Reports on Progress in Physics {\bf 72}, 126401 (2009).

\bibitem{Ped}  P. Pedri and L. Santos, Phys. Rev. Lett. {\bf 95}, 200404 (2005).
\bibitem{Cuev}  J. Cuevas, Boris A. Malomed, P. G. Kevrekidis, and D. J. Frantzeskakis, Phys. Rev. A {\bf 79}, 053608 (2009).
\bibitem {Bland} T. Bland, M. J. Edmonds, N. P. Proukakis, A. M. Martin, D. H. J. O'Dell, and N. G. Parker, Phys. Rev. A {\bf 92}, 063601 (2015).
\bibitem {Pawl} K. Pawlowski and K. Rzazewski, New J. Phys. {\bf 17}, 105006 (2015).
\bibitem {Edm}  M. J. Edmonds, T. Bland, D. H. J. O'Dell, and N. G. Parker, Phys. Rev. A {\bf 93}, 063617 (2016).
\bibitem{Lak}  K. Lakomy, R. Nath, and L. Santos, Phys. Rev. A {\bf 85}, 033618 (2012).
\bibitem{Baiz} B. B. Baizakov, S. M. Al-Marzoug, and H. Bahlouli, Phys. Rev. A {\bf 92}, 033605 (2015).

\bibitem{Abdul} F. Kh. Abdullaev and V. A. Brazhnyi, J. Phys. B: At. Mol. Opt. Phys. {\bf 45}, 085301 (2012).											
\bibitem {Sinh} S. Sinha and L. Santos,  Phys. Rev. Lett. {\bf 99}, 140406 (2007).
\bibitem {Yong} L.E. Young-s, P. Murunganandam and S.K. Adhikari, J. Phys. B: At. Mol. Opt. Phys. {\bf 44}, 101001 (2011).

\bibitem {Tik} I. Tikhonenkov, B. A. Malomed, and A. Vardi, Phys. Rev. A {\bf 78}, 043614 (2008).
\bibitem{Gli} G. Gligoric\', A. Maluckov, M. Stepic\', L. Hadžievski, and B. A. Malomed Phys. Rev. A {\bf 81}, 013633 (2010).
\bibitem {Ai} Ai-Xia Zhang and Ju-Kui Xue, Phys. Rev. A {\bf 82}, 013606 (2010).
\bibitem {Adh} S. K. Adhikari, P. Muruganandam,  J. Phys. B: At. Mol. Opt. Phys. {\bf 45}, 045301 (2012).
\bibitem{Fan} Z. Fan, Y. Shi, Y. Liu, W. Pang, Y. Li, B. A. Malomed, Phys. Rev. E {\bf 95}, 032226 (2017).
\bibitem {Adh1} S.K. Adhikari, Phys. Rev. A {\bf 89}, 043615 (2014); Phys. Rev. A {\bf 89}, 013630  (2014).

\bibitem {Tik1} I. Tikhonenkov, B. A. Malomed, and A. Vardi, Phys. Rev. Lett. {\bf 100}, 090406 (2008).
\bibitem{Eic} R. Eichler, D. Zajec, P. K\"oberle, J. Main, and G. Wunner, Phys. Rev. A {\bf 86}, 053611 (2012).
\bibitem{Kob}  P. K\"oberle, D. Zajec, G. Wunner, and B. A. Malomed, Phys.Rev. A {\bf 85}, 023630 (2012).
\bibitem{Rag} M. Raghunandan, C. Mishra, K. Lakomy, P. Pedri, L. Santos, and R. Nath,  Phys. Rev. A {\bf 92}, 013637 (2015).
\bibitem {Nath}  R. Nath, P. Pedri, and L. Santos, Phys. Rev. Lett. {\bf 101}, 210402 (2008).


\bibitem {Wang} D.-W. Wang, M. D. Lukin, and E. Demler, Phys. Rev. Lett. {\bf 97}, 180413 (2006).
\bibitem {Piko} A. Pikovski, M. Klawunn, G. V. Shlyapnikov, and L. Santos, Phys. Rev. Lett. {\bf 105}, 215302 (2010).
\bibitem {Misha} M. A. Baranov, A. Micheli, S. Ronen, and P. Zoller, Phys. Rev. A {\bf 83}, 043602 (2011).
\bibitem {Shi} S.-M. Shih and D.-W. Wang, Phys. Rev. A {\bf 79}, 065603 (2009).
\bibitem {Pot} A. C. Potter, E. Berg,D.-W.Wang, B. I. Halperin, and E. Demler, Phys. Rev. Lett. {\bf 105}, 220406 (2010).
\bibitem {Ros} M. Rosenkranz and W. Bao, Phys. Rev. A {\bf 84}, 050701(R) (2011).
\bibitem {Vol} A. G. Volosniev, D. V. Fedorov, A. S. Jensen, and N. T. Zinner, Phys. Rev. Lett. {\bf 106}, 250401 (2011).
\bibitem {Dalm}  M. Dalmonte, P. Zoller, G. Pupillo, Phys. Rev. Lett. {\bf 107}, 163202 (2011).
\bibitem{Santos1} R. Nath, P. Pedri, and L. Santos, Phys. Rev. A {\bf 76}, 013606 (2007).
\bibitem{Santos2} R. Nath, P. Pedri, and L. Santos, Phys. Rev. A {\bf 86}, 013610 (2012).
\bibitem{Chen}  Y.-A. Chen et {\it al.}, Nature Phys. {\bf 7},  61 (2011); S. Hofferberth et {\it al.}, Nature Phys. {\bf 2}, 710 (2006).

\bibitem{Fedo} A.K. Fedorov, S.I. Matveenko, V.I. Yudson, G.V. Shlyapnikov,  Sci. Rep. {\bf 6},  27448 (2016).
\bibitem{Boudj} A.Boudjem\^aa,  Phys. Lett. A {\bf 381}, 1745 (2017).
\bibitem{Boudj1} A.Boudjem\^aa,  J. Low. Temp. Phys {\bf 189}, 76 (2017).
\bibitem{Cai}  Y. Cai, M. Rosenkranz, Z. Lei, and W. Bao, Phys. Rev. A 82, 043623 (2010).
\bibitem{PitaevString} L. Pitaevskii and S. Stringari, Bose-Einstein Condensation, Oxford University Press (2003).
\bibitem{Boudj4} A.Boudjem\^aa, Phys. Rev. A {\bf 97}, 033627 (2018); Phys. Rev. A {\bf 98}, 033612 (2018).
\bibitem {Don} E. A. Donley, N. R. Claussen, S. L. Cornish, J. L. Roberts, E. A. Cornell, and C. E. Wieman, Nature {\bf 412}, 295 (2001).
\bibitem{Boudj2} A.Boudjem\^aa,  Commun. Nonlinear Sci. Numer. Simul. {\bf 33}, 85 (2016).
\bibitem{Boudj3}  A. Boudjemaa and G.V. Shlyapnikov, Phys. Rev. A {\bf 87}, 025601 (2013).

\bibitem{Nic}  De Nicholas Manton and Paul Sutcliffe, Topological Solitons, (Cambridge University Press 2004).





\end{thebibliography}
\end{document}